\documentclass[conference]{ifacconf}
\usepackage{amsmath, amssymb, amsfonts}
\usepackage{mathrsfs}
\DeclareMathOperator*{\argmax}{arg\,max}

\usepackage{algorithmic}
\usepackage{graphicx}
\usepackage{svg}
\usepackage{textcomp}
\usepackage{xcolor}
\usepackage{bm}
\usepackage{caption}
\usepackage{subcaption}
\usepackage{multirow}
\usepackage{tikz}
\usepackage{varwidth}
\usepackage{booktabs}
\usepackage{graphbox}
\let\labelindent\relax
\usepackage{enumitem}
\usepackage{natbib}
\usepackage{pdfpages}

\usetikzlibrary{positioning}
\usetikzlibrary{shapes.geometric, arrows}

\tikzstyle{box} = [rectangle, rounded corners, minimum height=1cm, text centered, draw=black]
\tikzstyle{arrow} = [thick,->,>=stealth]

\begin{document}
\begin{frontmatter}

\title{Relational Weight Optimization for Enhancing Team Performance in Multi-Agent Multi-Armed Bandits}

\author[first]{Monish Reddy Kotturu}
\author[second]{Saniya Vahedian Movahed}
\author[third]{Paul Robinette}
\author[fourth]{Kshitij Jerath}
\author[fifth]{Amanda Redlich}
\author[first]{Reza Azadeh}

\address[first]{Miner School of Computer \& Information Sciences, University of Massachusetts Lowell, Lowell, MA 01854}
\address[second]{University of Texas San Antonio, San Antonio, TX 78249}
\address[third]{Electrical \& Computer Engineering, University of Massachusetts Lowell, Lowell MA 01854}
\address[fourth]{Mechanical and Industrial Engineering, University of Massachusetts Lowell, Lowell MA 01854}
\address[fifth]{Mathematics \& Statistics, University of Massachusetts Lowell, Lowell MA 01854}

\thanks{This work was supported in part by  ARL W911NF20-2-0089 and NSF IIS-2112633.}

\begin{abstract}

We introduce an approach to improve team performance in a Multi-Agent Multi-Armed Bandit (MAMAB) framework using Fastest Mixing Markov Chain (FMMC) and Fastest Distributed Linear Averaging (FDLA) optimization algorithms. The multi-agent team is represented using a fixed relational network and simulated using the Coop-UCB2 algorithm. The edge weights of the communication network directly impact the time taken to reach distributed consensus. Our goal is to shrink the timescale on which the convergence of the consensus occurs to achieve optimal team performance and maximize reward. Through our experiments, we show that the convergence to team consensus occurs slightly faster in large constrained networks.
\end{abstract}

\end{frontmatter}

\section{Introduction}
Multi-Armed Bandits (MABs) are a class of reinforcement learning problems where an agent is presented with a set of arms (i.e., actions), with each arm giving a reward drawn from a probability distribution unknown to the agent \citep{lattimore2020bandit}. The goal of the agent is to maximize its total reward which requires balancing exploration and exploitation. MABs offer a simple model to simulate decision-making under uncertainty. Practical applications of MAB algorithms include news recommendations \citep{yang2018graph}, online ad placement \citep{aramayo2022multiarmed}, dynamic pricing \citep{babaioff2015dynamic}, and adaptive experimental design \citep{rafferty2019statistical}. In contrast to single-agent cases, in certain applications such as search and rescue, a team of agents should cooperate with each other to accomplish  goals by maximizing team performance. Such problems are solved using Multi-Agent Multi-Armed Bandit (MAMAB) algorithms \citep{xu2020collaborative}.
Most existing algorithms rely on the presence of multiple agents and try to solve the problem using the shared information among them ignoring the relationship between team members \citep{sankararaman2019social, rangi2018multi}. 

Few works, on the other hand, use graph representation \citep{shahrampour2017multi, madhushani2019heterogeneous} and grouping \citep{sankararaman2019social} to establish a specific team structure. Using a graph to represent the team behavior ensures that the relationship between the agents are held. However, existing works either do not consider the weight of each relationship (graph edges) \citep{madhushani2020dynamic, agarwal2021multi} or expect the user to manually set those weights \citep{moradipari2022collaborative}.

In this paper, we propose a new approach that combines graph optimization and MAMAB algorithms to enhance team performance by expediting the convergence to consensus of arm means. Our proposed approach:


\begin{itemize}[leftmargin=*]
    \item improves team performance by optimizing the edge weights in the graph representing the team structure in large constrained teams,
    \item does not require manual tuning of the graph weights,
    \item is independent of the MAMAB algorithm and only depends on the consensus formula, and
    \item formulates the problem as a convex optimization, which is computationally efficient for large teams.
\end{itemize}

We evaluate Coop-UCB2 combined with six different graph optimization approaches in various team structures, measured by time taken to converge to consensus of the best arm mean. Our results show that the proposed method outperforms existing graph-based MAMAB algorithms with manual tuning or other adjustment heuristics in large, constrained teams, but does not have a significant effect in small networks.

\section{Related Work}
\label{sec:rw}

Algorithms similar to Coop-UCB2 have been previously devised to simulate distributed learning in multi-agent networks. A distributed version of the Upper Confidence Bound (UCB) 1 algorithm \citep{shahrampour2017multi, auer2002finite} called d-UER was proposed that aims to minimize network regret while also using a network structure. It maintains the same principle as UCB in that it uses the upper confidence bound that relies on the network topology. Players vote for their estimated best arm iteratively and the network takes an action based on the majority vote. However, it focused on the case where the arms are dependent on the players. In \cite{sankararaman2019social}, the authors developed an algorithm in which agents choose to communicate with other agents uniformly and independently at random and only communicate the arm IDs. They show a significant reduction in the network regret in both full communication and no communication scenarios, mimicking different levels of collaboration. This and other works such as \cite{chawla2020gossiping, zhu2021federated, martinez2019decentralized} use gossip style algorithms to solve the MAMAB problem. However, the above algorithms do not optimize the consensus step of the communication process for faster convergence.



Literature on convex optimization methods involving distributed consensus models includes \textit{least-mean-square consensus} (LMSC) \citep{xiao2007distributed}, which takes fast linear iterations \citep{xiao2003fast} further by including the total mean-square deviation of each variable that converges to a steady-state value and minimizing this deviation. \cite{carli2007average} compared the behavior and performance of linear averaging algorithms for consensus problems in cases where transmission noise existed and where communication is quantized. They show that the algorithms fail to converge to a consensus when there is added white noise (with zero mean and bounded covariance), and they provide some solutions that involve discarding information from agents who are too close to the source of noise to prevent drift from initial average. \cite{zhou2013discrete} also examine the discrete average consensus problem with bounded noise and show that an increase upper bound for noise decreases the convergence accuracy.

\section{Background}
\label{sec:bg}

\subsection{Stochastic Multi-Armed Bandits}
A stochastic Multi-Armed Bandit (MAB) is a collection of probability distributions $\nu=(P_a: a\in \mathcal{A})$ over all the available arms (i.e., actions) $\mathcal{A}$. In each round, $t \in \{1, \dots, T\}$, the agent interacts with the MAB (i.e., environment) by selecting an arm $a(t) \in  \mathcal{A}$, and receives a reward $r(t) \in \mathbb{R}$ sampled from $P_{a(t)}$. Throughout this paper, we consider MABs from the Sub-Gaussian environment class $\mathcal{E}^k_{SG}(\sigma^2)=\{ P_a: \sigma-\textrm{Sub-Gaussian}\}$. For a $\sigma-$Sub-Gaussian random variable $X$, $\mathbb{P}(X \ge \epsilon) \le \exp(\frac{-\epsilon^2}{2\sigma^2})$ holds for any $\epsilon \ge 0$. Additionally, the tail of $X$ decays approximately as fast as that of $\mathcal{N}(0, \sigma^2)$ \citep{lattimore2020bandit}. The Gaussian, Bernoulli, and uniform distributions are examples of Sub-Gaussian distributions.

In single-agent MAB problems, the goal of the agent is to find a policy $\pi$ that maximizes the total reward $\sum_{t=1}^T r(t)$, where $T \in \mathbb{N}$ is the horizon. This is equivalent to minimizing the regret which is defined by $R_T(\pi, \nu)=T\mu^{*}(\nu) - \mathbb{E}[\sum_{t=1}^T r(t)]$, where $\mu^*=\argmax_{a \in \mathcal{A}} \mu_a(\nu)$ indicates the mean of the best arm and $\mu_a(\nu)$ is the mean of arm $a$ in the bandit $\nu$. The regret can be represented as $R_n(\pi, \nu)=\sum_{a \in \mathcal{A}} \Delta_a \mathbb{E}[n_a(T)]$, where $\Delta_a(\nu)=\mu^*(\nu)-\mu_a(\nu)$ is the immediate regret (i.e., the suboptimality gap) in the bandit $\nu$, and $n_a(\tau)=\sum_{t=1}^\tau \mathbb{I}\{a(t)=a\}$ is the total number of times action $a$ was selected after the end of round $\tau$.

\subsection{Multi-Agent Multi-Armed Bandits}
The goal of a team playing a Multi-Agent Multi-Armed Bandit is to maximize individual rewards over the horizon $T$, while playing the same bandit, cooperatively. In this setup, agents communicate with each other over a network, modeled by an undirected graph $\mathcal{G} = (V, E)$, where $V$ is the set of vertices representing $M$ agents and $E$ is the set of edges representing the connections between agents. Each agent, $k \in \{ 1, \ldots, M\}$, uses its current estimation to select an arm $A_t^k$ from the finite set of arms $\mathcal{A} = \{A_i \}$ for $i \in \{1, \ldots, N \}$, where $N \in \mathbb{N}$ and $|\mathcal{A}|=N$. The agent then receives a real value reward, updates its estimates, and shares the reward information of the selected arm with connected neighbors through $\mathcal{G}$. 
The objective of the cooperative multi-agent MAB problem is to maximize the expected total group reward that is equivalent to minimizing the expected group regret defined by $R_n(\pi, \nu)=\sum_{i=1}^N \sum_{k=1}^M \Delta_i \mathbb{E}[n_i^k(T)]$. 

In such a cooperative setting, agents update their estimations by combining their observations (i.e., realized rewards) and their neighbors' observations usually through a consensus process. The simplest and most common form of consensus, which is used in forming opinions in social learning, is the distributed averaging algorithm presented as $x(t+1)=P x(t)$, where the vector $x(t)$ represents agents' opinions in round $t$ and $P$ is the transition matrix in which $P_{ij}$ defines the weight that agent $i$ gives to agent $j$'s opinion. In such consensus process with no disturbances or new updates, all agents' opinions converge asymptotically to $x(0)$. The process of averaging consensus along its variants have been greatly studied in literature \citep{zhou2013discrete, boyd2004fastest, xiao2007distributed, xiao2003fast, carli2007average}. In the presence of an external signal, the distributed averaging consensus problem can be shown as $x(t+1)= P(x(t) + z(t))$, where $z(t)$ represents new updates, which in a multi-agent MAB problem is the vector of rewards realized by agents in round $t$.

\section{Methodology}
\label{sec:method}
\subsection{Problem Description}
The transition matrix, or Perron matrix, $P$ plays a significant role in the convergence behavior of the above-mentioned distributed averaging consensus process. In a cooperative multi-agent MAB algorithm, agents share the knowledge of observed reward information with their neighbors. 
The consensus process then updates the current estimate of the arms (i.e., actions) using the transition matrix $P$. Therefore, varying the values of the $P$ directly affects the performance of each agents and the team. In this paper, we aim at enhancing the multi-agent team performance by expediting the convergence to consensus. To achieve this goal, we propose an approach that combines graph optimization and multi-agent MAB algorithms. Given a relational network (i.e., the graph $\mathcal{G}$), our approach finds the optimal relational weights (i.e., elements of the transition matrix $P$) and uses those for the consensus process in the multi-agent MAB algorithm.




\subsection{Multi-Agent MAB Algorithm}
Our proposed method is independent of the choice of the multi-agent MAB algorithm and only relies on the use of distributed averaging consensus in the estimation of the arms' mean values through  shared realized reward between agents. Among existing algorithms, we focus on Coop-UCB2 \citep{landgren2020distributed} which is a multi-agent extension of the Upper Confidence Bound (UCB) algorithm \citep{auer2002finite} in which a group of agents perform cooperative decision making to solve a multi-armed bandit problem. Coop-UCB2 uses distributed consensus in a graph network structure for cooperative decision making. 
The agents communicate with each other through a network represented by a fixed undirected graph $\mathcal{G}=(V,E)$.
The algorithm starts by each agent sampling each arm once. In the next rounds, each agent selects the arm with maximum estimated mean according to the following estimation: 
\begin{equation}
    Q^k_i = \frac{\hat{s}^k_i}{\hat{n}^k_i} + \sigma_g \sqrt{\frac{2\gamma}{G(\eta)} \cdot \frac{\hat{n}^k_i + f(t - 1)}{M\hat{n}^k_i} \cdot \frac{\ln{(t - 1)}}{\hat{n}^k_i}},
    \label{equ:coopucb}
\end{equation}

\noindent where $Q^k_i$ is the estimation of the mean for arm $i$ generated by agent $k$, $\hat{s}^k_i$ is the estimation of the total reward given by arm $i$ until time-step $t$ for agent $k$, $\hat{n}^k_i$ is the number of times arm $i$ was selected by agent $k$ until time $t$, $f(t)$ is an increasing sub-logarithmic function of $t$ (e.g., $\sqrt{\log(t)}$), $\sigma_g \in \mathbb{R}_{+}$, $\gamma > 1$, $\eta \in (0,4)$, and $G(\eta) = 1 - \eta^2/16$ \citep{landgren2020distributed}.

Agents update their estimates cooperatively through a distributed consensus process represented using the following equations:
\begin{align}
    \hat{{n}}_i(t) &= P(\hat{{n}}_i(t - 1) + {\xi}_i(t)),\label{equ:consensus1} \\ 
    \hat{{s}}_i(t) &= P(\hat{{s}}_i(t - 1) + {r}_i(t)),
    \label{equ:consensus2}
\end{align}
\noindent where ${\xi}_i(t)$ is a vector representing the number of times the action $i$ was selected and ${r}_i(t)$ is the reward vector at time-step $t$. In Coop-UCB2, $P$ is defined as a row stochastic matrix (also known as Perron matrix) obtained from the following equation:
\begin{equation}
    P = I_M - \frac{\kappa}{d_{max}}L,
    \label{equ:landgren_step}
\end{equation}
\noindent where $I_M$ is the identity matrix of order $M$ (the number of agents), $\kappa \in (0, 1]$ is a step size parameter, and $d_{max} = max\{deg(i) | i \in \{1, ..., M\}\}$. $L$ is the Laplacian matrix calculated from the adjacency matrix $A = [A_{ij}]$ for graph $\mathcal{G}$ as
\begin{equation}
    L_{ij} = \left\{ \begin{array}{ll}
    \sum^M_{k=1, k \ne i} A_{ik}, & \ j = i \\
    -A_{ij}, & \ j \ne i
    \end{array} \right..
\end{equation}

However, this method of calculating the edge weights of $P$ in \eqref{equ:landgren_step} is independent of the relational network's structure and complexity. Consequently, using this methods does not result in the optimal team performance in many situations. We consider a team performance optimal if the network of agents converges to the optimal consensus as fast as possible.



\subsection{Optimization using Fastest Mixing Markov Chain}
With the goal of discovering an optimal $P$ matrix, we consider formulating the problem in the context of Markov chain on the given graph. We can define a Markov chain over the undirected graph $\mathcal{G}=(V,E)$ representing our multi-agent team. Our goal is to optimize the weight $P_{ij}$ of each edge $(i,j) \in E$, which is the probability of transitioning from node $i$ to node $j$. The assigned probabilities must be non-negative and for each node, the sum of the probabilities of links connected to the node must equal to one. In other words, $P_{ij}=p(x(t+1)=i | x(t)=j)$ for $i,j=1,...,M$, where $x(t) \in \{1,...,M\}$, for $t \in \mathbb{Z}_+$ represents states. The transition matrix must satisfy $P_{ij} \ge 0 $, $\textbf{1}^\top P = \textbf{1}^\top$, $P=P^\top$, and $P_{ij}=0$ for $(i,j) \notin E$, where $\textbf{1}$ is the vector of all ones. The second largest eigenvalue of $P$ is called the mixing rate and is defined as $\max \{ \lambda_2, -\lambda_n\}$, where $1=\lambda_1 \ge \lambda_2 \ge ... \ge \lambda_n$ in which $\lambda_i$s are the eigenvalues of $P$. The smaller mixing rate results in faster mixing in the Markov chain towards the equilibrium distribution $(1/M)\textbf{1}$. The Fastest Mixing Markov Chain (FMMC) is the optimization problem to find the minimum mixing rate. By applying the idea of FMMC to our consensus process, we can obtain optimal relational network weights that expedite the convergence. 

The optimization problem can be represented as the following convex form: 
\begin{equation}
    \begin{array}{ll}
        \text{minimize} & \|P-(1/M)\textbf{1}\textbf{1}^\top \|_2 \\
        \text{subject to} & P \ge 0, P\textbf{1} = \textbf{1}, P = P^\top, \\
        & P_{ij} = 0, (i, j) \not \in E,
    \end{array}
    \label{eq:fmmc}
\end{equation}
\noindent where $\|.\|_2$ represents the spectral norm.

The formulation in \eqref{eq:fmmc} can be expressed as a Semi Definite Program (SDP) \citep{alizadeh1995interior, boyd2004convex} by introducing a scalar slack variable $s \in \mathbb{R}$ as follows:
\begin{equation}
    \begin{array}{ll}
        \text{minimize} & s \\
        \text{subject to} & -sI \preceq P-(1/M)\textbf{1}\textbf{1}^\top \preceq sI, \\
        & P \ge 0, P\textbf{1} = \textbf{1}, P = P^\top, \\
        & P_{ij} = 0, (i, j) \not \in E,
    \end{array}
    \label{equ:fmmc_s}
\end{equation}
\noindent where the symbol $\preceq$ denotes matrix inequality. The optimization problem in \eqref{equ:fmmc_s} can be solved using interior-point algorithms for SDP \citep{alizadeh1995interior}.





\subsection{Optimization using Fastest Distributed Linear \\ Averaging}
Alternatively, a distributed consensus over a network of agents can be optimized by finding a linear iteration that is able to calculate the average weights of nodes and edges in the given graph \citep{xiao2003fast}. Fastest Linear Distributed Linear Averaging (FDLA) achieves this with a set of constraints on $P$ to solve the problem faster. The FDLA optimization problem can be formulated as
\begin{equation}
    \begin{array}{ll}
        \text{minimize} & \rho(P - (1/M)\textbf{11}^T), \\
        \text{subject to} & P \in \mathscr{S}, P = P^T, P\textbf{1} = \textbf{1},
    \end{array}
    \label{equ:fdla}
\end{equation}

\noindent where $\rho$ is the spectral radius of $P$ and \\ $\mathscr{S} = \left\{ P \in \mathbb{R}^{M \times M} | P_{ij} = 0 \text{ if } {i, j} \not\in E \text{ and } i \ne j \right\}$.

The formulation in \eqref{equ:fdla} can be expressed as an SDP by introducing a scalar slack variable $s \in \mathbb{R}$ as follows:
\begin{equation}
    \begin{array}{ll}
        \text{minimize} & s \\
        \text{subject to} & -sI \preceq P-(1/M)\textbf{1}\textbf{1}^\top \preceq sI, \\
        & P \in \mathscr{S}, P\textbf{1} = \textbf{1}, P = P^\top.
    \end{array}
    \label{equ:fdla_s}
\end{equation}

Similar to FMMC, the optimization problem in \eqref{equ:fdla_s} can be solved using interior-point algorithms for SDP \citep{alizadeh1995interior}. Unlike FMMC, FDLA, however, allows for negative values in $P$. It has to be noted that although negative values are not obvious choices for weights that represent the relationship between the agents, they can result in faster convergence in certain situations.

\subsection{On the Convergence of the Running Consensus}
Our running average consensus works according to $x(t+1)=P(x(t)+z(t))$, where $\textbf{1}^\top P = \textbf{1}^\top$, $P \textbf{1} = \textbf{1}$, i.e., $P$ is doubly stochastic, and $z(t) \sim \mathcal{N}(\mu, \sigma_2)$ is external observation (i.e. rewards) obtained by playing an arm with the true mean $\mu$, and variance $\sigma^2$. If $z(t)=0$, our formula reduces to the discrete average consensus and converges to $\bar{x} = \frac{1}{M}\sum_{i=1}^M x(0) = \frac{\textbf{1}\textbf{1}\top}{M}x(0)$. When $z(t) \sim \mathcal{N}(0, \sigma^2)$ (i.e., white noise), it has been shown that the consensus system is stable \citep{carli2007average, hatano2005agreement}. Here, however, we assume that $z(t)$ is the observation of the agent received after taking an action. So, it can be written:
\begin{align}
    x(t) &= P^t x(0) + \sum_{k=1}^t z(t-k+1)P^k \\ \nonumber
    &= P^t x(0) + \sum_{k=1}^t \mu P^k + \sum_{k=1}^t \varepsilon(t-k+1)P^k
\end{align}

\noindent where $\varepsilon$ represents the noise. We know that $\lim_{t\xrightarrow{} \infty} P^t x(0) = \bar{x} \textbf{1}$. Since $P$ is doubly stochastic with the maximum eigenvalue equal to one, it can be shown that $\lim_{t\xrightarrow[]{}\infty} \sum_{k=1}^t z(t-k+1)P^k = C\textbf{1}$, where $C$ is a constant. Consequently, we conclude that the running consensus converges to a constant $\lim_{t\xrightarrow{} \infty} x(t) = \bar{x}\textbf{1} + C\textbf{1} = C'\textbf{1}$.

\begin{figure}
    \centering
    \begin{subfigure}[b]{0.3\linewidth}
        \centering\captionsetup{justification=centering}
        \includegraphics[width=\linewidth]{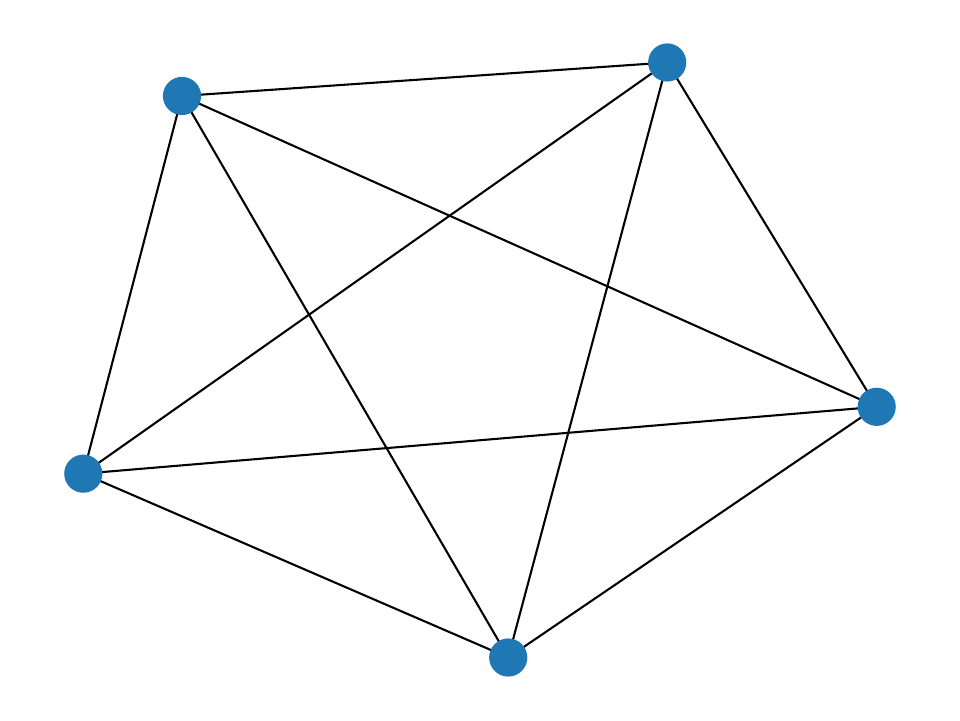}
        \label{fig:all-to-all-network}
    \end{subfigure}
    \centering
    \begin{subfigure}[b]{0.3\linewidth}
        \centering\captionsetup{justification=centering}
        \includegraphics[width=\linewidth]{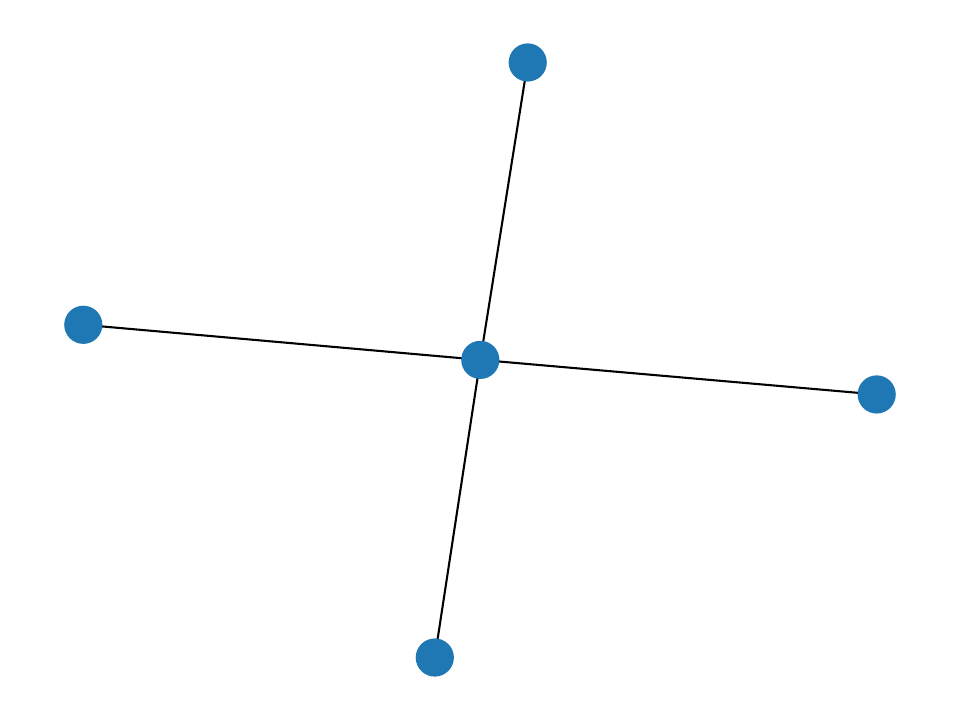}
        \label{fig:star_network}
    \end{subfigure}
    \centering
    \begin{subfigure}[b]{0.3\linewidth}
        \centering\captionsetup{justification=centering}
        \includegraphics[width=\linewidth]{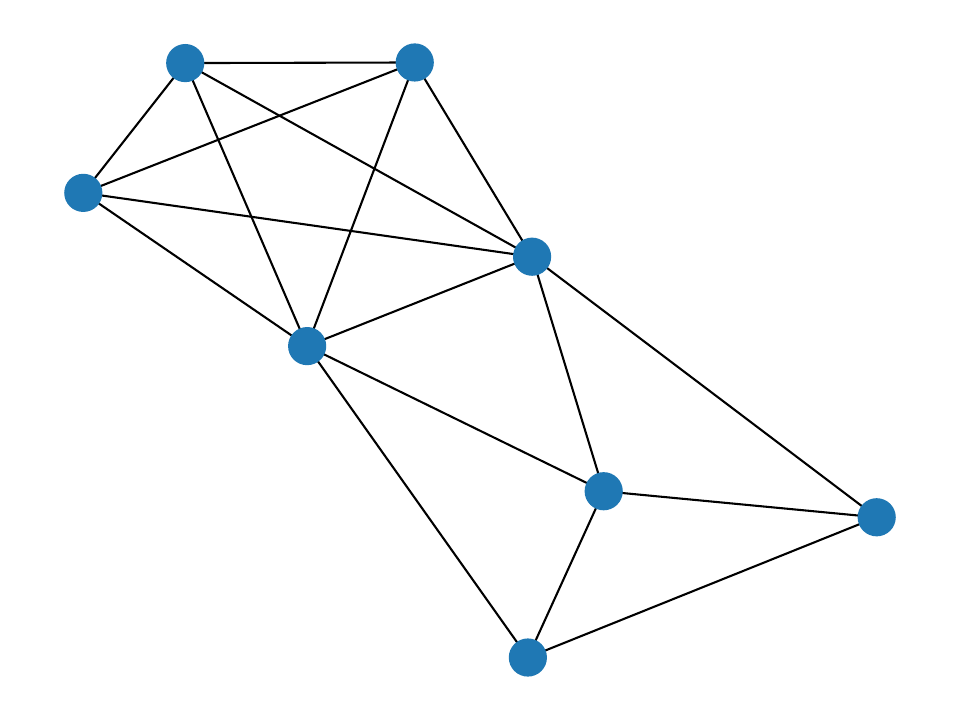}
        \label{fig:8-agents_network}
    \end{subfigure}
    \centering
    \begin{subfigure}[b]{0.3\linewidth}
        \centering\captionsetup{justification=centering}
        \includegraphics[width=\linewidth]{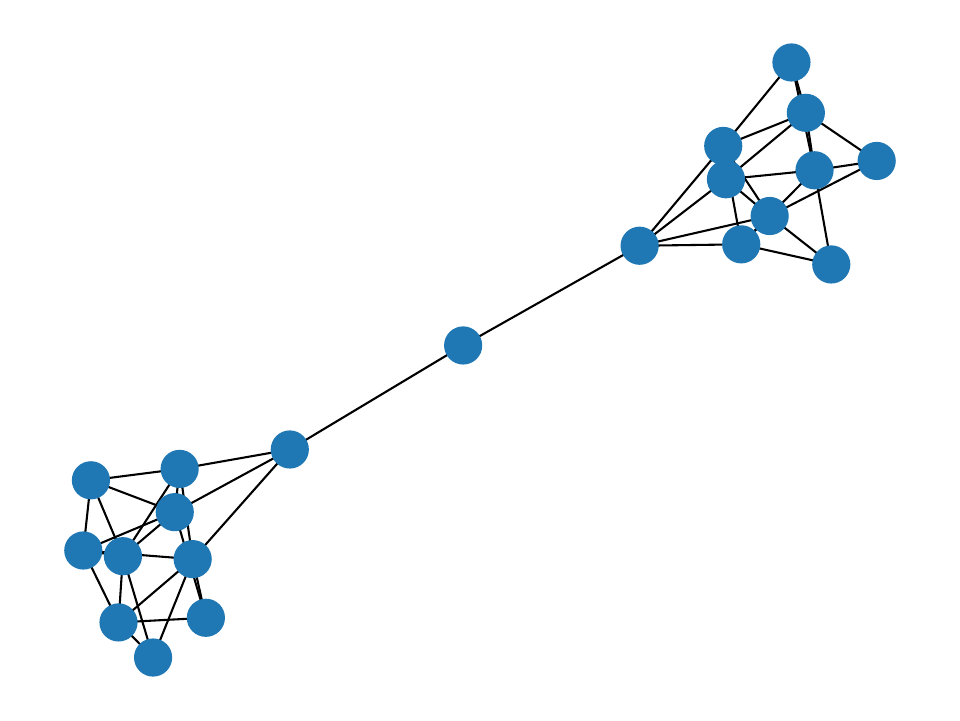}
        \label{fig:2-clusters_network}
    \end{subfigure}
    \centering
    \begin{subfigure}[b]{0.3\linewidth}
        \centering\captionsetup{justification=centering}
        \includegraphics[width=\linewidth]{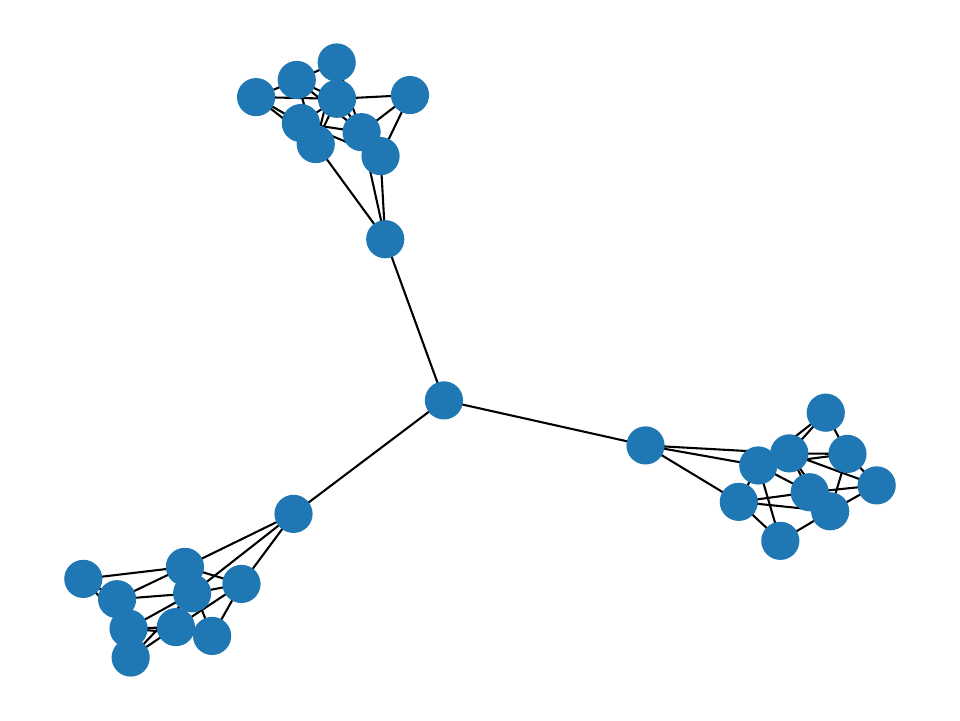}
        \label{fig:3-clusters_network}
    \end{subfigure}
    \centering
    \begin{subfigure}[b]{0.3\linewidth}
        \centering\captionsetup{justification=centering}
        \includegraphics[width=\linewidth]{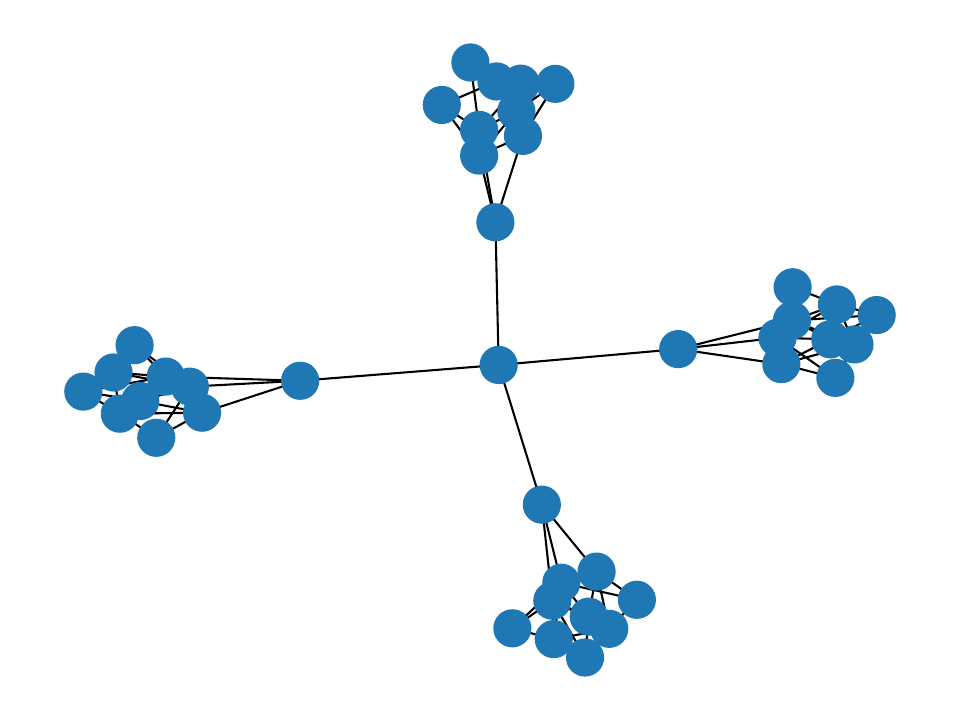}
        \label{fig:4-clusters_network}
    \end{subfigure}
    \centering
    \caption{Graphical representation of the networks}
    \label{fig:networks_figures}
\end{figure}

\begin{figure*}[ht]
    \centering
    \begin{subfigure}[b]{0.8\linewidth}
        \centering\captionsetup{justification=centering}
        \includegraphics[width=\linewidth]{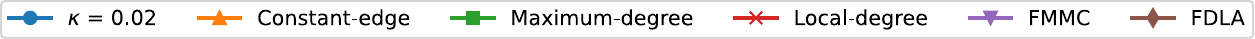}
        \label{fig:legend}
    \end{subfigure}
    \begin{subfigure}[]{\linewidth}
        \centering
        \begin{subfigure}[b]{0.3\linewidth}
            \centering\captionsetup{justification=centering}
            \includegraphics[width=\linewidth]{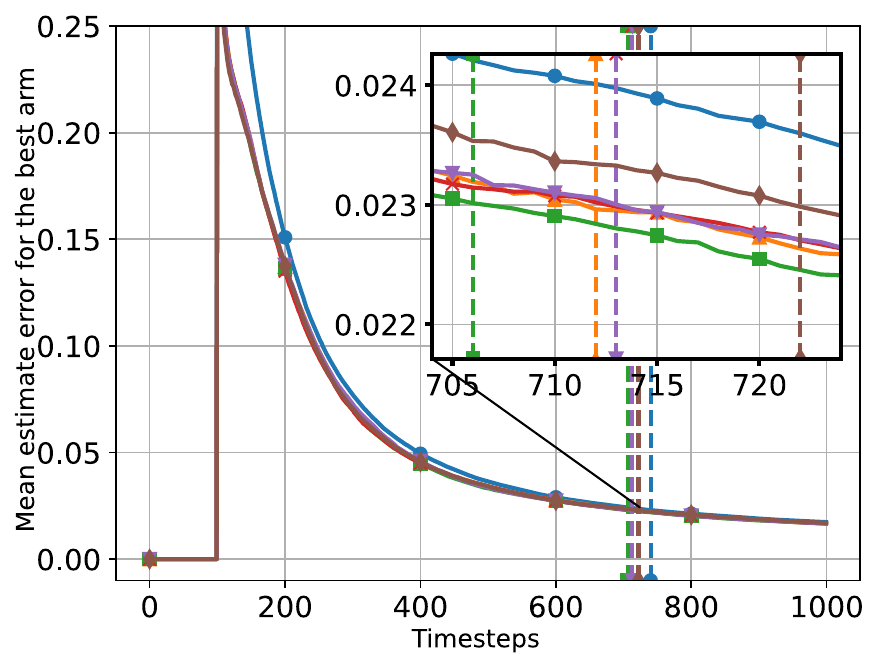}
            \caption{All-to-All}
            \label{fig:errorcomparison_all_to_all}
        \end{subfigure}
        \begin{subfigure}[b]{0.3\linewidth}
            \centering\captionsetup{justification=centering}
            \includegraphics[width=\linewidth]{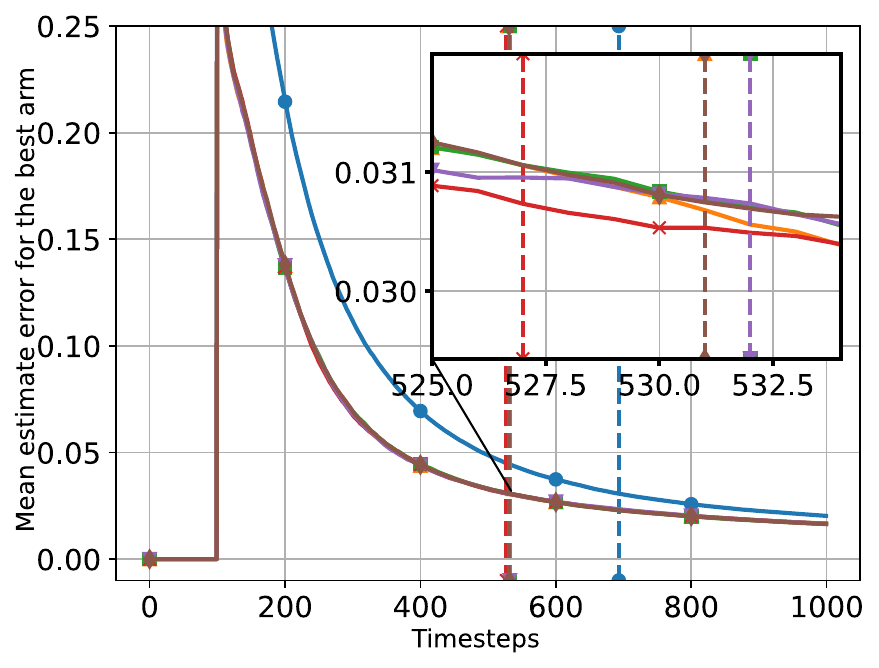}
            \caption{Star}
            \label{fig:errorcomparison_star}
        \end{subfigure}
        \begin{subfigure}[b]{0.3\linewidth}
            \centering\captionsetup{justification=centering}
            \includegraphics[width=\linewidth]{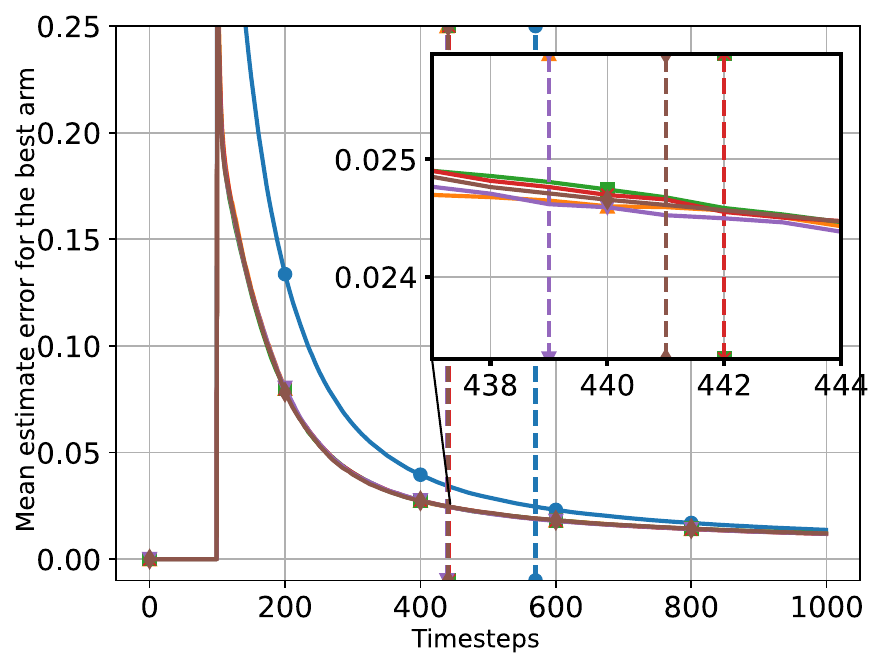}
            \caption{8-agents}
            \label{fig:errorcomparison_8_agents}
        \end{subfigure}
        \begin{subfigure}[b]{0.3\linewidth}
            \centering\captionsetup{justification=centering}
            \includegraphics[width=\linewidth]{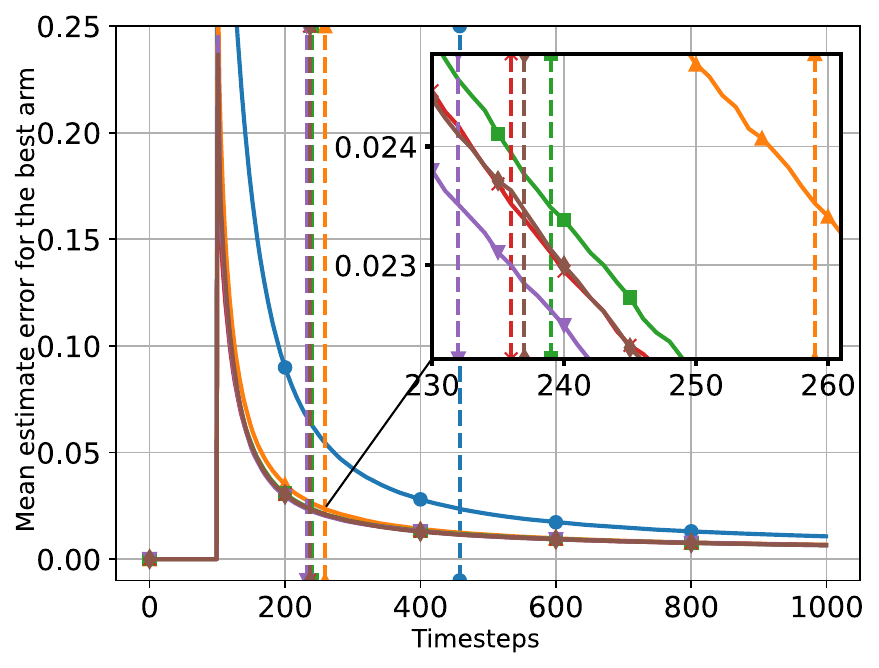}
            \caption{2 Clusters}
            \label{fig:errorcomparison_2_clusters}
        \end{subfigure}
        \begin{subfigure}[b]{0.3\linewidth}
            \centering\captionsetup{justification=centering}
            \includegraphics[width=\linewidth]{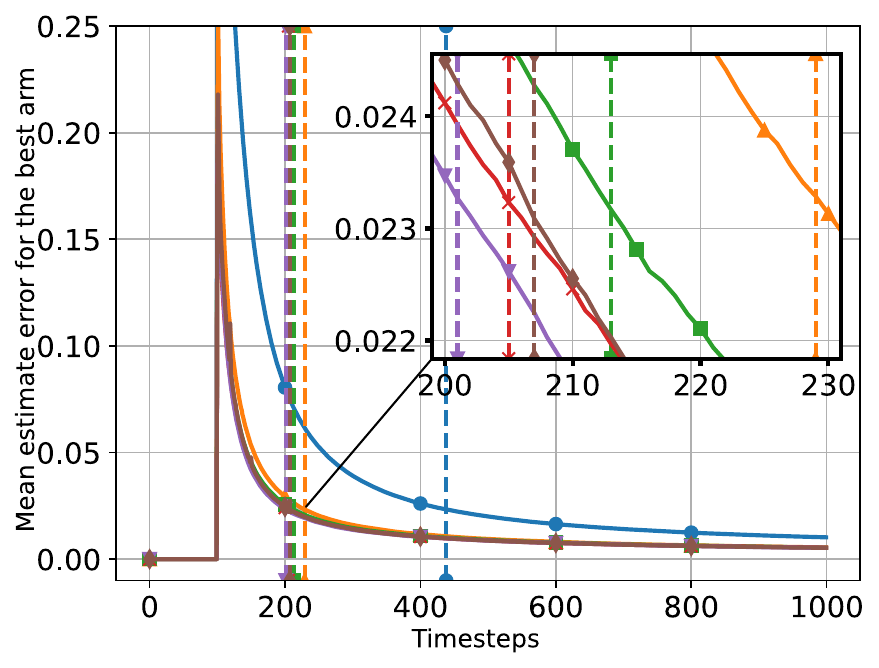}
            \caption{3 Clusters}
            \label{fig:errorcomparison_3_clusters}
        \end{subfigure}
        \begin{subfigure}[b]{0.3\linewidth}
            \centering\captionsetup{justification=centering}
            \includegraphics[width=\linewidth]{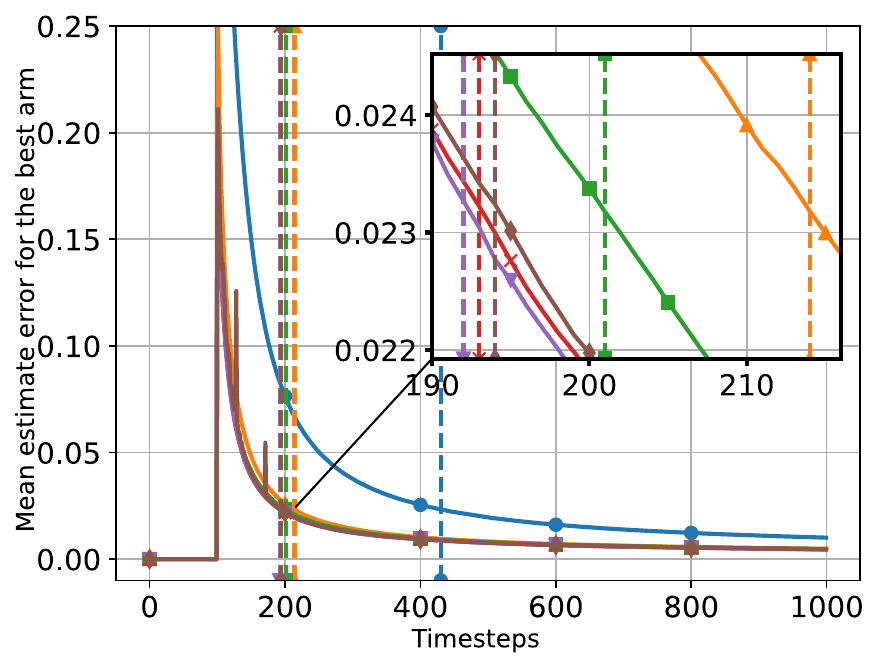}
            \caption{4 Clusters}
            \label{fig:errorcomparison_4_clusters}
        \end{subfigure}
    \end{subfigure}
    \caption{\small{Comparison of team average of the errors between the estimated and true means of the best arm in different networks. The vertical dashed lines represent the time taken by the network to reach 5\% of the final value of the largest error among all algorithms.}}
    \label{fig:charts}
\end{figure*}

\begin{table*}[ht]
    \centering
    \caption{\small{Convergence factors in networks with weights optimized using various methods.}}
    \begin{tabular}{ccccccccccccccc}
        \toprule
        \multicolumn{1}{c}{Edge weight} & \multicolumn{2}{c}{All-to-All} & \multicolumn{2}{c}{Star} & \multicolumn{2}{c}{8-agents} & \multicolumn{2}{c}{2-Cluster} & \multicolumn{2}{c}{3-Cluster} & \multicolumn{2}{c}{4-Cluster} \\
        \cmidrule(lr){2-3} \cmidrule(lr){4-5} \cmidrule(lr){6-7} \cmidrule(lr){8-9} \cmidrule(lr){10-11} \cmidrule(lr){12-13}
        \multicolumn{1}{c}{generation method} & $\rho$ & $\tau$ & $\rho$ & $\tau$ & $\rho$ & $\tau$ & $\rho$ & $\tau$ & $\rho$ & $\tau$ & $\rho$ & $\tau$ \\
        \midrule
        $\kappa = 0.02$ & 0.975            & 39.498         & 0.995          & 199.5          & 0.995          & 196.5          & 0.999          & 3838.7          & 0.999          & 3950.2          & 0.999          & 3952.5          \\
        Constant-edge   & \textbf{3.3e-16} & \textbf{0.028} & \textbf{0.667} & \textbf{2.466} & 0.655          & 2.363          & 0.981          & 52.011          & 0.982          & 53.899          & 0.982          & 54.155          \\
        Maximum-degree  & 0.250            & 0.721          & 0.750          & 3.476          & 0.746          & 3.416          & 0.987          & 76.283          & 0.987          & 78.513          & 0.987          & 78.559          \\
        Local-degree    & 0.250            & 0.721          & 0.750          & 3.476          & 0.743          & 3.369          & 0.984          & 60.277          & 0.983          & 57.623          & 0.983          & 57.668          \\
        FMMC            & 5.2e-08          & 0.060          & 0.750          & 3.476          & 0.667          & 2.466          & 0.974          & 37.718          & 0.977          & 43.724          & 0.981          & 52.279          \\
        FDLA            & 2.9e-08          & 0.058          & \textbf{0.667} & \textbf{2.466} & \textbf{0.600} & \textbf{1.958} & \textbf{0.969} & \textbf{31.983} & \textbf{0.974} & \textbf{37.653} & \textbf{0.977} & \textbf{42.667} \\
        \bottomrule
    \end{tabular}
    \label{table:rho_comp}
\end{table*}

\section{Experiments}

\subsection{Experimental Setups}

Our experiments consist of 100-armed bandits where for each arm $a$, the true mean $\mu_a$ is sampled from $\mathcal{N}(0, 1)$. For each arm $a$, the rewards are sampled from $\mathcal{N}(\mu_a, \sigma^2)$, where $\sigma_a = 1$ unless mentioned otherwise. Each experiment involve 10,000 runs each of which includes playing a different bandit for 1,000 time steps. The networks we use in our experiments are shown in Fig. \ref{fig:networks_figures}. They include 5-agent networks adopted from \cite{landgren2020distributed}, namely, all-to-all and star networks, and an 8-agent network adopted from \cite{xiao2003fast}. We also introduce our own set of networks that are created by joining small clusters by a common ``parent" node. This is to increase the complexity of the network by decreasing the closeness centrality between nodes from different clusters and increasing the betweenness centrality of the parent node and thus, introducing a constraint on how fast the information is propagated throughout the team. In each experiment, we performed the FMMC and FDLA optimization on the given graph as presented in Section~\ref{sec:method} and used the Coop-UCB2 algorithm in \eqref{equ:coopucb} with the distributed consensus in \eqref{equ:consensus1} and \eqref{equ:consensus2}. 

\subsection{Heuristics}
Here, we introduce a few heuristics that could be used for computing an optimal $P$ that guarantees convergence of distributed linear averaging iteration.

\noindent \textbf{Constant-edge weights:} A heuristic that achieves this goal is \textit{constant-edge weights} that simply sets $P = I_M - \alpha L$. In this method, all the self weights add up to 1 ($P\textbf{1} = \textbf{1}$) and the edge weights are equal to $\alpha$.

\noindent \textbf{Maximum-degree weights:} another heuristic that follows the same principle, but uses $\alpha = 1/d_{max}$, where $d_{max}$ is the maximum degree in the graph. It is proven that distributed linear averaging converges using the constant-edge weights method where $\alpha \in (0, 1/d_{max})$ \citep{xiao2003fast}.

\noindent \textbf{Local-degree weights:} Another heuristic that generates $P$ by assigning the weights based on the maximum of the two neighboring nodes as $P_{ij} = \frac{1}{max\{d_i, d_j\}}$, where $d_i$ and $d_j$ are the degrees of nodes $i$ and $j$ in the graph, respectively.

\subsection{Evaluation Metrics}

We evaluate the performance of our optimized weight matrix $P$ using two metrics: the spectral radius and the convergence time.

\noindent \textbf{Spectral radius ($\rho$):} represents the  asymptotic convergence factor defined as
\begin{equation}
    \rho = \sup_{x(0) \ne \bar{x}} \lim_{t \to \infty} \left( \frac{\|x(t)-\bar{x}\|_2}{\|x(0)-\bar{x}\|_2} \right)^{1/t}.
\end{equation}

\noindent \textbf{Convergence time ($\tau$):}  defined as
\begin{equation}
    \tau = \frac{1}{log(1/\rho)}.
\end{equation}

Assuming $*$ indicates the best arm in a given MAB, we compare the average estimated mean for the best arm among all above-mentioned methods defined as $\frac{\hat{s}^*}{\hat{n}^*}$, where $\hat{s}^*$ and $\hat{n}^*$ are estimation of the total reward given by the best arm and the number of times the best arm was selected, respectively. We define the team average of the errors between the estimated and true means as $\delta$:
\begin{align}
    \delta = \frac{1}{M} \cdot \sum^{M}_{k=1}\left( \frac{\hat{s}^*_k}{\hat{n}^*_k} - \mu^* \right).
\end{align}




\subsection{Results and Research Insights}
\label{sec:results}
All teams optimized using FMMC and FDLA theoretically perform better in terms of $\rho_\textrm{asym}$ and $\tau_\textrm{asym}$ as reported in Table \ref{table:rho_comp}. However, in terms of $\delta$, the smaller 5 and 8-agent teams perform inconsistently between the heuristics, FMMC, and FDLA. The heuristics perform better in every case as represented by the vertical dashed lines, which signify the time taken to reach 5\% of the largest error among all heuristic methods and algorithms.

As the network complexity increases when transitioning to the clustered networks, we see a trend where FMMC does consistently better than most of the heuristics, usually closely followed by local-degree or FDLA. Even if $\rho$ and $\tau$ show that FDLA should be the most optimal, we hypothesize that it does not translate to an actual performance improvement due to negative weights produced by FDLA. Although not shown in the figures, it should be noted that a network with the same number of nodes and edges as the clustered networks but without the constraint of having one parent node do not have the same improvement when optimized using FMMC and FDLA. This maybe due to this large randomly generated graph still having similar properties to the smaller networks that affect communication, such as high closeness centralities among all nodes. Therefore, we think that the problem and team size play a significant role in the effectiveness of optimization.


While networks optimized with both FMMC and FDLA performed better than others in the larger networks, it is important to consider the differences between FMMC and FDLA. FMMC ensures that the resulting transition matrix $P$ has positive weights as its optimization problem imposes the constraint that $P \ge 0$. On the other hand, FDLA does not impose the same constraint. So, the resulting matrix may contain negative weights, which may be important to consider depending on the application of this network optimization approach.

\section{Conclusions}

In this paper, we propose a method that uses consensus optimization methods to optimize edge weights of a communication graph and improve the team performance of a multi-agent network playing a MAB. The results of our experiments show that this optimization step can improve the team performance in terms of the time taken to reach a consensus on the best arm in large constrained networks, which consist of several small clusters connected by a single parent node. In our future work, we would like to consider real-world networks with different properties that may also benefit from optimized edge weights.

\bibliography{ifacconf}

\begin{thebibliography}{26}
\providecommand{\natexlab}[1]{#1}
\providecommand{\url}[1]{\texttt{#1}}
\providecommand{\urlprefix}{URL }
\expandafter\ifx\csname urlstyle\endcsname\relax
  \providecommand{\doi}[1]{doi:\discretionary{}{}{}#1}\else
  \providecommand{\doi}{doi:\discretionary{}{}{}\begingroup \urlstyle{rm}\Url}\fi

\bibitem[{Agarwal et~al.(2021)Agarwal, Aggarwal, and Azizzadenesheli}]{agarwal2021multi}
Agarwal, M., Aggarwal, V., and Azizzadenesheli, K. (2021).
\newblock Multi-agent multi-armed bandits with limited communication.
\newblock \emph{arXiv preprint arXiv:2102.08462}.

\bibitem[{Alizadeh(1995)}]{alizadeh1995interior}
Alizadeh, F. (1995).
\newblock Interior point methods in semidefinite programming with applications to combinatorial optimization.
\newblock \emph{SIAM journal on Optimization}, 5(1), 13--51.

\bibitem[{Aramayo et~al.(2022)Aramayo, Schiappacasse, and Goic}]{aramayo2022multiarmed}
Aramayo, N., Schiappacasse, M., and Goic, M. (2022).
\newblock A multiarmed bandit approach for house ads recommendations.
\newblock \emph{Marketing Science}.

\bibitem[{Auer et~al.(2002)Auer, Cesa-Bianchi, and Fischer}]{auer2002finite}
Auer, P., Cesa-Bianchi, N., and Fischer, P. (2002).
\newblock Finite-time analysis of the multiarmed bandit problem.
\newblock \emph{Machine learning}, 47(2), 235--256.

\bibitem[{Babaioff et~al.(2015)Babaioff, Dughmi, Kleinberg, and Slivkins}]{babaioff2015dynamic}
Babaioff, M., Dughmi, S., Kleinberg, R., and Slivkins, A. (2015).
\newblock Dynamic pricing with limited supply.

\bibitem[{Boyd et~al.(2004{\natexlab{a}})Boyd, Boyd, and Vandenberghe}]{boyd2004convex}
Boyd, S., Boyd, S.P., and Vandenberghe, L. (2004{\natexlab{a}}).
\newblock \emph{Convex optimization}.
\newblock Cambridge university press.

\bibitem[{Boyd et~al.(2004{\natexlab{b}})Boyd, Diaconis, and Xiao}]{boyd2004fastest}
Boyd, S., Diaconis, P., and Xiao, L. (2004{\natexlab{b}}).
\newblock Fastest mixing markov chain on a graph.
\newblock \emph{SIAM Review}, 46(4), 667--689.
\newblock \doi{10.1137/S0036144503423264}.
\newblock \urlprefix\url{https://doi.org/10.1137/S0036144503423264}.

\bibitem[{Carli et~al.(2007)Carli, Fagnani, Frasca, Taylor, and Zampieri}]{carli2007average}
Carli, R., Fagnani, F., Frasca, P., Taylor, T., and Zampieri, S. (2007).
\newblock Average consensus on networks with transmission noise or quantization.
\newblock In \emph{2007 European Control Conference (ECC)}, 1852--1857.
\newblock \doi{10.23919/ECC.2007.7068829}.

\bibitem[{Chawla et~al.(2020)Chawla, Sankararaman, Ganesh, and Shakkottai}]{chawla2020gossiping}
Chawla, R., Sankararaman, A., Ganesh, A., and Shakkottai, S. (2020).
\newblock The gossiping insert-eliminate algorithm for multi-agent bandits.
\newblock In \emph{International conference on artificial intelligence and statistics}, 3471--3481. PMLR.

\bibitem[{Hatano et~al.(2005)Hatano, Das, and Mesbahi}]{hatano2005agreement}
Hatano, Y., Das, A.K., and Mesbahi, M. (2005).
\newblock Agreement in presence of noise: pseudogradients on random geometric networks.
\newblock In \emph{Proceedings of the 44th IEEE Conference on Decision and Control}, 6382--6387. IEEE.

\bibitem[{Landgren et~al.(2021)Landgren, Srivastava, and Leonard}]{landgren2020distributed}
Landgren, P., Srivastava, V., and Leonard, N. (2021).
\newblock Distributed cooperative decision making in multi-agent multi-armed bandits.
\newblock \emph{Automatica}, 125, 109445.
\newblock \doi{10.1016/j.automatica.2020.109445}.

\bibitem[{Lattimore and Szepesv{\'a}ri(2020)}]{lattimore2020bandit}
Lattimore, T. and Szepesv{\'a}ri, C. (2020).
\newblock \emph{Bandit algorithms}.
\newblock Cambridge University Press.

\bibitem[{Madhushani and Leonard(2019)}]{madhushani2019heterogeneous}
Madhushani, U. and Leonard, N.E. (2019).
\newblock Heterogeneous stochastic interactions for multiple agents in a multi-armed bandit problem.
\newblock In \emph{2019 18th European Control Conference (ECC)}, 3502--3507. IEEE.

\bibitem[{Madhushani and Leonard(2020)}]{madhushani2020dynamic}
Madhushani, U. and Leonard, N.E. (2020).
\newblock A dynamic observation strategy for multi-agent multi-armed problem.
\newblock In \emph{2020 European Control Conference (ECC)}, 1677--1682. IEEE.

\bibitem[{Mart{\'\i}nez-Rubio et~al.(2019)Mart{\'\i}nez-Rubio, Kanade, and Rebeschini}]{martinez2019decentralized}
Mart{\'\i}nez-Rubio, D., Kanade, V., and Rebeschini, P. (2019).
\newblock Decentralized cooperative stochastic bandits.
\newblock \emph{Advances in Neural Information Processing Systems}, 32.

\bibitem[{Moradipari et~al.(2022)Moradipari, Ghavamzadeh, and Mahnoosh}]{moradipari2022collaborative}
Moradipari, A., Ghavamzadeh, M., and Mahnoosh, A. (2022).
\newblock Collaborative multi-agent stochastic linear bandits.
\newblock In \emph{2022 American Control Conference (ACC)}, 2761--2766. IEEE.

\bibitem[{Rafferty et~al.(2019)Rafferty, Ying, Williams et~al.}]{rafferty2019statistical}
Rafferty, A., Ying, H., Williams, J., et~al. (2019).
\newblock Statistical consequences of using multi-armed bandits to conduct adaptive educational experiments.
\newblock \emph{Journal of Educational Data Mining}, 11(1), 47--79.

\bibitem[{Rangi and Franceschetti(2018)}]{rangi2018multi}
Rangi, A. and Franceschetti, M. (2018).
\newblock Multi-armed bandit algorithms for crowdsourcing systems with online estimation of workers' ability.
\newblock In \emph{Proceedings of the 17th International Conference on Autonomous Agents and MultiAgent Systems}, 1345--1352.

\bibitem[{Sankararaman et~al.(2019)Sankararaman, Ganesh, and Shakkottai}]{sankararaman2019social}
Sankararaman, A., Ganesh, A., and Shakkottai, S. (2019).
\newblock Social learning in multi agent multi armed bandits.
\newblock 3(3).
\newblock \doi{10.1145/3366701}.
\newblock \urlprefix\url{https://doi.org/10.1145/3366701}.

\bibitem[{Shahrampour et~al.(2017)Shahrampour, Rakhlin, and Jadbabaie}]{shahrampour2017multi}
Shahrampour, S., Rakhlin, A., and Jadbabaie, A. (2017).
\newblock Multi-armed bandits in multi-agent networks.
\newblock In \emph{2017 IEEE International Conference on Acoustics, Speech and Signal Processing (ICASSP)}, 2786--2790.
\newblock \doi{10.1109/ICASSP.2017.7952664}.

\bibitem[{Xiao and Boyd(2003)}]{xiao2003fast}
Xiao, L. and Boyd, S. (2003).
\newblock Fast linear iterations for distributed averaging.
\newblock In \emph{42nd IEEE International Conference on Decision and Control (IEEE Cat. No.03CH37475)}, volume~5, 4997--5002 Vol.5.
\newblock \doi{10.1109/CDC.2003.1272421}.

\bibitem[{Xiao et~al.(2007)Xiao, Boyd, and Kim}]{xiao2007distributed}
Xiao, L., Boyd, S.P., and Kim, S.J. (2007).
\newblock Distributed average consensus with least-mean-square deviation.
\newblock \emph{J. Parallel Distributed Comput.}, 67, 33--46.

\bibitem[{Xu et~al.(2020)Xu, Tao, and Shen}]{xu2020collaborative}
Xu, X., Tao, M., and Shen, C. (2020).
\newblock Collaborative multi-agent multi-armed bandit learning for small-cell caching.
\newblock \emph{IEEE Transactions on Wireless Communications}, 19(4), 2570--2585.

\bibitem[{Yang and Toni(2018)}]{yang2018graph}
Yang, K. and Toni, L. (2018).
\newblock Graph-based recommendation system.
\newblock In \emph{2018 IEEE Global Conference on Signal and Information Processing (GlobalSIP)}, 798--802. IEEE.

\bibitem[{Zhou et~al.(2013)Zhou, He, Cheng, and Chen}]{zhou2013discrete}
Zhou, M., He, J., Cheng, P., and Chen, J. (2013).
\newblock Discrete average consensus with bounded noise.
\newblock In \emph{52nd IEEE Conference on Decision and Control}, 5270--5275.
\newblock \doi{10.1109/CDC.2013.6760718}.

\bibitem[{Zhu et~al.(2021)Zhu, Zhu, Liu, and Liu}]{zhu2021federated}
Zhu, Z., Zhu, J., Liu, J., and Liu, Y. (2021).
\newblock Federated bandit: A gossiping approach.
\newblock \emph{Proceedings of the ACM on Measurement and Analysis of Computing Systems}, 5(1), 1--29.

\end{thebibliography}

\end{document}